\documentclass{elsart}
\usepackage{amsfonts,amsmath}
\newcommand{\ket}[1]{|#1\rangle}
\newcommand{\bra}[1]{\langle #1|}

\newcommand{\Tr}{\text{Tr}}
\begin{document}
\begin{frontmatter}
\title{Off-diagonal quantum holonomy along density operators}
\author[filipp]{Stefan Filipp\corauthref{fi}} and
\ead{sfilipp@ati.ac.at}
\corauth[fi]{Corresponding author.}
\author[sj]{Erik Sj\"{o}qvist}
\ead{erik.sjoqvist@kvac.uu.se}
\address[filipp]{Atominstitut der {\"{O}}sterreichischen 
Universit{\"{a}}ten, Stadionallee 2, A-1020 Vienna, Austria; \\ 
Institut Laue Langevin, Bo\^ite Postale 156, F-38042 Grenoble
Cedex 9, France}
\address[sj]{Department of Quantum Chemistry, Uppsala University,
Box 518, Se-751 20 Uppsala, Sweden}

\date{\today}
\begin{abstract}
Uhlmann's concept of quantum holonomy for paths of density operators
is generalised to the off-diagonal case providing insight into the
geometry of 
state space when the Uhlmann holonomy is undefined. Comparison
with previous off-diagonal geometric phase definitions is carried out and an example comprising the
transport of a Bell-state mixture is given. 
\end{abstract}
\end{frontmatter}

\section{Introduction}
Investigations in the polarisation of light have led Pancharatnam
\cite{pancharatnam56} in the mid-fifties to a notion of relative
phase. Some thirty years later Berry \cite{berry84} shed new light on
the relative phase between two states by introducing a decomposition
into a dynamical and a geometric part in the special case of
adiabatically guided Hamiltonians. The geometric phase depends wholly
on the shape of the curve in the subjacent parameter space
representing the evolution of the system. These new findings
encouraged further investigations leading towards more general
notions, in particular the spectrum of Berry's discovery has soon been
enlarged to include nonadiabatic \cite{aharonov87}, noncyclic, and
nonunitary \cite{samuel88} evolutions.  Furthermore, a kinematic
derivation has been given by Mukunda and Simon \cite{mukunda93}, and
Pati defined the geometric phase via a reference section
\cite{pati95a,pati95b}. Since the special case of pure states in
quantum mechanics is not sufficient for dealing with realistic
problems, it is a legitimate question whether one can ascribe a
geometric phase to mixed states. Uhlmann
\cite{uhlmann86,uhlmann93,uhlmann96} proposed a phase holonomy for
paths of density operators utilizing a purification scheme of mixed
into pure states obtained via a certain parallelity condition. 
Based on Uhlmann's definition there has been further investigations
 in  the parallel transport of density matrices in mixed state space
 relating also to differential geometric techniques 
\cite{hubner93,dittmann-rudolph92,uhlmann95,corach01}.
The experimental testability of the Uhlmann holonomy has been
addressed recently in \cite{tidstrom03}. Another approach taken by
Chaturvedi et al. \cite{chaturvedi03} uses methods from differential
geometry to obtain a mixed state geometric phase. In addition,
Sj\"{o}qvist et al. \cite{sjoqvist00} generalised the geometric phase
to mixed states starting from an interferometry setup. This latter
phase concept has recently been tested experimentally in nuclear
magnetic resonance \cite{du03} and single photon interferometry
\cite{ericsson05}.  The two approaches in
Refs. \cite{uhlmann86,sjoqvist00} have been found out to be
incompatible for arbitrary nondegenerate mixed states
\cite{slater02,ericsson03a} fully coinciding only in the pure state
limit.

From Pancharatnam's relative phase concept it is easily deducible
that one runs into problems while trying to calculate the relative
phase between orthogonal states. This defect has been cured by
Manini and Pistolesi \cite{manini00} (see also \cite{mukunda01}
for a general framework in terms of Bargmann invariants) for pure
states and by the current authors for mixed states
\cite{filipp03a} (see also Refs. \cite{filipp03b,tong04}) by
introducing off-diagonal geometric phases that may be well-defined
when the phases discussed in Refs.
\cite{pancharatnam56,samuel88,sjoqvist00} are not. In this paper,
we follow this line of reasoning first in spotting nodal points of
the Uhlmann quantum holonomy, i.e., configurations where the
relative phase factor between the initial and final density
operator turns out to be undefined, and second in proposing
generalised holonomy quantities that may be well-defined in such
situations. Furthermore we give examples to show the relevance of
this extension.

\section{Quantum holonomy and its nodal points}
\label{sec:uhlmannholonomy}

Let ${\mathcal C}: t \in [0,\tau] \mapsto \rho(t)$ be a path of
density operators. A lift of ${\mathcal C}$ is a path
$\widetilde{\mathcal C}: t \in [0,\tau] \mapsto W(t)$ such that $\rho
(t) = W(t) W^\dagger (t)$ with the amplitude $W(t) = \rho^{1/2}(t)
V(t)$, the one-parameter family of partial isometries $V(t)$ being the
phase factors of $W(t)$ along the path $\widetilde{\mathcal C}$. If
for any pair of amplitudes along the path the following condition is
fulfilled
\begin{equation}
W^\dagger(t) W(t') > 0,
\end{equation}
then $W(t)$ and  $W(t')$ are said to be parallel. For infinitesimally
close $t$ and $t'$ the parallelity condition reads
\begin{equation}
\label{eq:infparallel}
W^\dagger(t) d W(t) = d W^\dagger(t) W(t) > 0,
\end{equation}
which, if valid for all $t\in [0,\tau]$, defines a parallel lift of
the path ${\mathcal C}$.

For an initial amplitude $W(0) = \rho^{1/2}(0)$ with the phase factor
$V(0)$ chosen to be the identity operator on the support of $\rho(0)$,
the final amplitude $W(\tau)$ is given by $W(\tau) = \rho^{1/2}(\tau)
\widetilde{V} (\tau)$, if $W(t)$ fulfils the parallel transport
condition in Eq. (\ref{eq:infparallel}).
A gauge invariant quantity of the path ${\mathcal C}$ is given by the
operator $W(\tau) W^\dagger (0)$ \cite{uhlmann93} that can be written
in terms of the initial and final state as
\begin{equation}
\label{eq:invariant}
W(\tau) W^\dagger (0) = \rho^{1/2}(\tau) \widetilde{V}(\tau)
\rho^{1/2}(0),
\end{equation}
which is only dependent on the path of states. That $W(\tau)
W^\dagger(0)$ and the relative phase factor $\widetilde{V}(\tau)$
restricted by the support of the initial density operator are indeed
gauge invariant can be seen from the fact that a local gauge
transformation $W(t) \mapsto W'(t) = W(t)S(t)$ by a partial isometry
$S(t)$ has already been fixed by imposing the parallel transport
condition on the path $\widetilde{\mathcal{C}}$
\cite{uhlmann93}. Moreover, the remaining gauge freedom by a
$t$-independent partial isometry $S$ along the path, i.e., $W(t)
\mapsto W'(t) = W(t) S$ leaves Eq. (\ref{eq:invariant})
invariant. Note that the Uhlmann quantum holonomy invariant $W(\tau)
W^\dagger (0)$ is well-defined if the path $\mathcal{C}$ is
sufficiently regular, in particular, the support of the density
operators involved should change smoothly
\cite{uhlmann93}.

By introducing the functional
\begin{equation}
\label{eq:nuberry}
\nu_{\mathcal{C}}(A) \equiv \arg \Tr[A W(\tau) W^\dagger(0)]
\end{equation}
on the observables $A\in \mathcal{B}(\mathcal{H})$, where
$\mathcal{B}(\mathcal{H})$ denotes the algebra of linear operators
acting on the Hilbert space $\mathcal{H}$ under consideration, we
obtain the well-known Berry phase \cite{berry84,aharonov87,samuel88}
for pure states as follows.

Consider the initial state $\ket{0}\in \mathcal{H}$ and its
standard purification $W(0) = \ket{0}\bra{a}\in
\mathcal{H}\otimes\mathcal{H}^*$, where $\bra{a}$ can be any
element of the dual Hilbert space $\mathcal{H}^*$ and the unitary
evolution operator $U(t)\in \mathcal{B}(\mathcal{H})$. Then the
amplitude at time $t$ is given by $W(t) = U(t) \ket{0}\bra{a}$.
The Uhlmann holonomy invariant $W(\tau) W^\dagger(0)$ is then
$U(\tau)\ket{0}\bra{0}$ and by setting $A = 1$ in Eq.
(\ref{eq:nuberry}) we obtain the usual geometric phase
$\nu_{\mathcal{C}}(1) = \arg \bra{0}U(\tau)\ket{0}$ for $U(t)$
fulfilling the parallel transport condition $\bra{0} U^{\dagger}
(t) \dot{U} (t) \ket{0} = 0$.

In the pure state limit above, $\nu_{\mathcal{C}} (1)$ has nodal
points (i.e., is undefined) for orthogonal initial and final
states. In general, let ${\mathcal H}_1 \oplus {\mathcal H}_2 
\oplus \ldots \oplus {\mathcal H}_n$ be an orthogonal sum 
decomposition of the Hilbert space $\mathcal{H}$ of the system, 
where $n \leq \dim {\mathcal H}$ with equality for the case 
discussed by Manini and Pistolesi \cite{manini00}. Then, if the 
initial state $\rho(0)$ with support in ${\mathcal H}_k$ ($\rho(0) 
\in {\mathcal B}({\mathcal H}_k)$) evolves to the state
$\rho(\tau)$ with support in ${\mathcal H}_l$ ($l\neq k$),
$\nu_{\mathcal{C}} (1)$ is undefined since the trace vanishes as is
apparent from Eq. (\ref{eq:invariant}). If this happens $\rho(0)$ and
$\rho(\tau)$ are said to be orthogonal.  Analogous to the mixed state
generalization \cite{filipp03a} of the off-diagonal geometric phase
defined in \cite{manini00} based on the interferometric approach
to mixed state geometric phases taken in \cite{sjoqvist00} an
off-diagonal quantum holonomy based on Uhlmann's idea \cite{uhlmann86}
of a relative phase factor accompanying the parallel transport of
mixed states can be constructed to cover such situations.

\section{Off-diagonal generalisation of quantum holonomies}
\label{sec:definition1}
The criteria imposed on such a generalising quantity are: (a) the
reducability to the off-diagonal pure state geometric phase, (b) the
inclusion of Uhlmann's relative phase as a special instance, in the
same manner as the geometric phase for mixed states in
\cite{sjoqvist00} is included in its generalisation
\cite{filipp03a}, (c) the invariance under a gauge transformation 
$\rho_k \mapsto \rho_k'= S_k\rho_k S_k^\dagger$ of the initial 
mixed states $\rho_k$ by partial isometries $S_k$ on the support 
of $\rho_k$, and (d) it should potentially be well-defined for 
orthogonal initial and final states.

The latter is evidently the main motivation in constructing an
off-diagonal quantum holonomy for mixed states in order to obtain
geometric information whenever the Uhlmann holonomy shows nodal
points.  To this end we assume a set of initial density operators
$\rho_k(0), \ k=1,\ldots,n,$ each of which with support in the
corresponding Hilbert space $\mathcal{H}_k$. For a given evolution one
obtains the final set of density operators $\rho_k(\tau),
k=1,\ldots,n,$ and one can associate a path $\mathcal{C}_k$ to each
$\rho_k$ giving rise to the holonomies $\mathcal{V}_k(\mathcal{C}_k)$.
Instead of regarding the latter separately one can consider gauge
invariant quantities depending on some or all of the paths revealing
insight into the geometric properties of state space in case of one or
more vanishing $\mathcal{V}_k(\mathcal{C}_k)$'s.

As in Uhlmanns original construction the basic elements are the
purifications $W_k(0)$ and $W_k(\tau)$ of the initial and the final
states represented by $\rho_k(0)$ and $\rho_k(\tau)$ and their
Hermitian conjugates to guarantee the reduction to Uhlmann's
definition. The only issue left is the correct order to ensure the
validity of criteria (a)-(d) from above. As we will see there is
essentially just one possibility.

First of all, we note that there are two alternatives stated in the
literature for Uhlmann's holonomies, either $W(\tau)W^\dagger (0)$
\cite{uhlmann93,uhlmann96} or $W(0) W^\dagger (\tau)$ \cite{uhlmann86},
both being invariants depending only on the path $\mathcal{C}: t \in
[0,\tau] \mapsto \rho(t)$. We adopt the former since it reduces to
Pancharatnam's phase factor for pure states while the latter entails a
phase with opposite sign.  As for the ordering ensuring the reducibility
to the pure state case the amplitudes of the initial and final states
have to appear next to each other, i.e., the off-diagonal quantum holonomy
is of the form $W_1(\tau)W_1^\dagger (0) W_2(\tau) W_2^\dagger (0)\ldots
W_k(\tau)W_k^\dagger (0)$.  

We are now ready to define the \emph{off-diagonal quantum holonomy
invariants} as
\begin{eqnarray}
\label{eq:definitionholonomy1}
 & & \mathcal{X}_{j_1 \ldots j_l}^{(l)} [\mathcal{C}_{j_1} \ldots
\mathcal{C}_{j_l}] 
\nonumber \\ 
 & & \equiv W_{j_1}(\tau) W_{j_1}^\dagger(0)
W_{j_2}(\tau) W_{j_2}^\dagger(0) \ldots W_{j_l}(\tau)
W_{j_l}^\dagger (0)
\nonumber \\
 & & = \mathcal{X}_{j_1}^{(1)} [\mathcal{C}_{j_1}]
\mathcal{X}_{j_2}^{(1)} [\mathcal{C}_{j_2}] \ldots
\mathcal{X}_{j_l}^{(1)} [\mathcal{C}_{j_l}] ,
\end{eqnarray}
where $l=1,\ldots,n$. Evidently, $\mathcal{X}_{j_k}^{(1)}
[\mathcal{C}_{j_k}]$, $k=1,\ldots,n,$ is the Uhlmann
holonomy invariant for the path ${\mathcal C}_{j_k} : t\in [0,\tau]
\rightarrow \rho_{j_k} (t)$ and each $\mathcal{X}_{j_k}
[\mathcal{C}_{j_k}]$ comprises a relative phase factor
$\widetilde{V}_{j_k}$ depending only on the path
$\mathcal{C}_{j_k}$.
Clearly, this definition comprises Uhlmann's holonomy invariant
for $l=1$ fulfilling therefore criterion (b).

$\mathcal{X}\equiv \mathcal{X}_{j_1 \ldots j_l}^{(l)}
[\mathcal{C}_{j_1} \ldots \mathcal{C}_{j_l}]$ can be decomposed
either as $\mathcal{X} = (\mathcal{X} \mathcal{X}^\dagger)^{1/2}
\mathcal{U}_{\textrm{\tiny{R}}}$ (right polar decomposition) or as
$\mathcal{X} = \mathcal{U}_{\textrm{\tiny{L}}} (
\mathcal{X}^\dagger \mathcal{X} )^{1/2}$ (left polar
decomposition), where $\mathcal{U}_{\textrm{\tiny{R}}}$ and
$\mathcal{U}_{\textrm{\tiny{L}}}$ are partial isometries on the
right or left support of $\mathcal{X}$ (denoted by
$\textrm{r-supp\,} \mathcal{X}$ or $\textrm{l-supp\,}
\mathcal{X}$, respectively). The polar decomposition theorem
\cite[p. 197]{reed80} states that
$\mathcal{U}_{\textrm{\tiny{L}}}$ is unique under the condition
that $\textrm{Ker\,} \mathcal{U}_{\textrm{\tiny{L}}} =
\textrm{Ker\,} \mathcal{X}$. Furthermore, $\textrm{l-supp\,}
\mathcal{U}_{\textrm{\tiny{L}}} = \textrm{l-supp\,} \mathcal{X}$.
For $\mathcal{U}_{\textrm{\tiny{R}}}$ the uniqueness condition is
$\textrm{l-supp\,} \mathcal{U}_{\textrm{\tiny{R}}} =
\textrm{l-supp\,} \mathcal{X}$ and by using these restrictions of
the supports of $\mathcal{U}_{\textrm{\tiny{R,L}}}$ we can show
the equality $\mathcal{U}_{\textrm{\tiny{L}}}=
\mathcal{U}_{\textrm{\tiny{R}}}$: Inserting the projection
operator
$\mathcal{U}_{\textrm{\tiny{R}}}\mathcal{U}_{\textrm{\tiny{R}}}^\dagger$
onto the left support of $\mathcal{U}_{\textrm{\tiny{R}}}$ (which
is equal to the left support of $\mathcal{X}$) into the right
polar decomposition of $\mathcal{X}$ we obtain 
\begin{equation}
\mathcal{X}  =  (\mathcal{X} \mathcal{X}^\dagger)^{1/2}
\mathcal{U}_{\textrm{\tiny{R}}} =
\mathcal{U}_{\textrm{\tiny{R}}}\mathcal{U}_{\textrm{\tiny{R}}}^\dagger
(\mathcal{X} \mathcal{X}^\dagger)^{1/2}
\mathcal{U}_{\textrm{\tiny{R}}} 
=  \mathcal{U}_{\textrm{\tiny{L}}}
(\mathcal{X}^\dagger \mathcal{X})^{1/2} 
\end{equation}

The last equality follows from the uniqueness of the polar
decomposition and therefore we have
\begin{equation}
\mathcal{U}_{\textrm{\tiny{L}}} =
\mathcal{U}_{\textrm{\tiny{R}}}=\mathcal{U}. \end{equation} The
holonomy $\mathcal{U}$ is the Uhlmann analogue to the off-diagonal
geometric phase factors defined in \cite{manini00} for pure
states and in \cite{filipp03a} for mixed states.

We shall now consider the nodal points of $\mathcal{X}$. To this end
we note that the nodal points of $\mathcal{X}_1^{(1)}[\mathcal{C}_1]$
has been located for vanishing $\nu_{\mathcal{C}}(1)$ in Section
\ref{sec:uhlmannholonomy}. Therefore, we introduce the generalised
functional
\begin{eqnarray} 
\label{eq:generalberryfunctional}
\nu_{\mathcal{C}_{j_1}\ldots \mathcal{C}_{j_l}}^{(l)}(A) & \equiv & 
\arg \Tr \big[ A W_{j_1}(\tau) W_{j_1}^\dagger(0) 
 W_{j_2}(\tau)
W_{j_2}^\dagger(0) \ldots W_{j_l}(\tau)W_{j_l}^\dagger (0)\big]
\nonumber \\
 & = & \arg \Tr \big[ A\mathcal{X} \big] .
\end{eqnarray}
Trivially, $\nu_{\mathcal{C}_{j_1}\ldots
\mathcal{C}_{j_l}}^{(l)}(1)$ is undefined for $\mathcal{X}=0$. For
$\mathcal{X}\neq 0$ a sufficient condition for a nodal point of
$\nu_{\mathcal{C}_{j_1}\ldots \mathcal{C}_{j_l}}^{(l)}(1)$ is
orthogonal supports of the positive Hermitian parts of the left
and right polar decomposition of $\mathcal{X}$. This can be seen
by noting at first that the left and right support of the operator
$\mathcal{X}$ is given by the support of $\mathcal{X}
\mathcal{X}^\dagger$ and $\mathcal{X}^\dagger \mathcal{X}$,
respectively, and in addition that the trace of $\mathcal{X}$
vanishes for nonoverlapping left and right support. Since
$\mathcal{X} \mathcal{X}^\dagger$ and $\mathcal{X}^\dagger
\mathcal{X}$ appear in the positive Hermitian parts of the polar
decomposition the nodal points of $\nu_{\mathcal{C}_{j_1}\ldots
\mathcal{C}_{j_l}}^{(l)}(1)$ are necessary for orthogonal left and
right supports of $\mathcal{X}$.

Let us have a detailed look at the right and left support of
$\mathcal{X}$. The left support is given by
\begin{eqnarray}
\mathcal{X}\mathcal{X}^\dagger & = & 
\rho_{j_1}^{1/2}(\tau)
\widetilde{V}_{j_1}(\tau) \rho_{j_1}^{1/2}(0) 
\nonumber \\ 
 & & \times \rho_{j_2}^{1/2}(\tau)
\widetilde{V}_{j_2}(\tau) \ldots \rho_{j_l}^{1/2}(\tau)
\widetilde{V}_{j_l}(\tau)
\nonumber \\
 & & \times \rho_{j_l} (0) \widetilde{V}_{j_l}^\dagger (\tau) \ldots
\rho_{j_1}^{1/2}(0) \widetilde{V}_{j_1}^\dagger (\tau)
\rho_{j_1}^{1/2}(\tau)
\end{eqnarray}
and the right support
\begin{eqnarray}
\mathcal{X}^\dagger\mathcal{X} & = & 
\rho_{j_l}^{1/2}(0) \widetilde{V}_{j_l}^\dagger (\tau)
\rho_{j_l}^{1/2} (\tau) \ldots
\rho_{j_1}^{1/2}(0) 
\nonumber \\
 & & \times \widetilde{V}_{j_1}^\dagger (\tau) \rho_{j_1}(\tau) 
\widetilde{V}_{j_1}(\tau) \rho_{j_1}^{1/2}(0)
\nonumber \\
 & & \times \rho_{j_2}^{1/2}(\tau) \widetilde{V}_{j_2}(\tau) \ldots
\rho_{j_l}^{1/2} (\tau) \widetilde{V}_{j_l} \rho_{j_l}^{1/2}(0).
\end{eqnarray}
These are apparently only orthogonal in the case where 
$\rho_{j_1}(\tau)$ and $\rho_{j_l}(0)$ have orthogonal support and
this in turn can be avoided by a proper choice of initial states.
These choices of the $\rho_{j_k}(0),\ k = 1,\ldots,l$ are evidently
not unique, one can take any state $\rho_{j_l} \in
\mathcal{B}(\mathcal{H}_{j_l})$ for a given
$\rho_{j_1}$ with the minimal requirement that the $\rho_{j_k}(\tau)$
have overlapping supports at least with $\rho_{j_{k-1}}(0)$ where
the indices $k$ have to be considered modulo $n$. This is
equivalent to nonvanishing transition probability from
$\rho_{j_k}(\tau)$ to $\rho_{j_{k-1}}(0)$ \cite{uhlmann76}.

To assure that the off-diagonal quantum holonomy invariants
$\mathcal{X}_{j_1 \ldots j_l}^{(l)} [\mathcal{C}_{j_1} \ldots
\mathcal{C}_{j_l}]$ fulfil all necessary criteria, we note that
the $\mathcal{X}_{j_1 \ldots j_l}^{(l)} [\mathcal{C}_{j_1} \ldots
\mathcal{C}_{j_l}]$'s are only dependent on the paths
$\mathcal{C}_{j_k}$ by the same reasoning as for the $l=1$ case.
In fact, the final amplitude $W_{j_k}(t)$ of each initial state
$\rho_{j_k}(0)$ is determined by the parallel transport condition
in Eq. (\ref{eq:infparallel}) up to a $t$-independent partial
isometry $S$. This latter global gauge leaves $\mathcal{X}_{j_1
\ldots j_l}^{(l)} [\mathcal{C}_{j_1} \ldots \mathcal{C}_{j_l}]$
invariant even for distinct choices $S=S_{j_k}$ for the different
constituent initial states showing the validity of criterion (c).
There is no need then to state a prescription for the relations
between the $\rho_k$'s \cite{remark1} and the reference states can be
chosen arbitrarily taking into account that their supports belong to
the correct subspaces.

We now rewrite the parallel transport mechanism in the particular
case of mixed states undergoing unitary evolution. The standard
purification of a mixed state $\rho(0) = \sum \lambda_j \ket{\psi_j}
\bra{\psi_j}$ with $\sum_j \lambda_j = 1$ and $\ket{\psi_j}$ being
a basis diagonalising $\rho(0)$ is $W(0) = \sum_j \sqrt{\lambda_j}
\ket{\psi_j}\bra{\phi_j}$, i.e., ${\mathcal H}$ is extended by an
ancilla Hilbert space ${\mathcal H}' = {\mathcal H}^{\ast}$, where the
$\bra{\phi_j} \in {\mathcal H}^{\ast}$ form a basis in the ancilla
part. Subjected to the unitary evolution $\rho(0)\mapsto \rho(t) =
U(t)\rho(0)U^\dagger(t), \ t\in [0,\tau]$, the path of the
purifications $t\mapsto W(t)$ has to fulfil the parallelity condition
(\ref{eq:infparallel}). This latter path can be described by applying
a partial isometry $B(t) \in \mathcal{B}(\mathcal{H}')$ resulting in
\begin{equation}
\label{eq:Wpath}
W(t) = U(t) \rho^{1/2}(0) B(t),
\end{equation}
where $B(t) = U^\dagger(t) V(t)$ and $U(t)$ are related via
the parallel transport condition Eq. (\ref{eq:infparallel}).
Inserting (\ref{eq:Wpath}) into (\ref{eq:infparallel}) we obtain 
\begin{equation}
\label{eq:paralleltransportoperators}
2\rho^{1/2}(0) U^\dagger(t)\dot{U}(t) \rho^{1/2}(0) = 
B(t)\dot{B}^\dagger(t) \rho(0)  - \rho(0) \dot{B}(t) B^\dagger(t) ,
\end{equation}
where the dot denotes the derivative with respect to the parameter
$t$. If $\rho(0)$ is pure, $\rho(0) = \ket{\psi_j}\bra{\phi_j}$, Eq.
(\ref{eq:paralleltransportoperators}) simplifies to
\begin{equation}
\label{eq:pureparallelity}
\bra{\psi_j}{U^{\dagger}(t) \dot{U}(t)} \ket{\psi_j} = 
\bra{\phi_j}{B^{\dagger}(t) \dot{B}(t)} \ket{\phi_j}.
\end{equation}

To verify that Eq. (\ref{eq:definitionholonomy1}) is consistent
with known results we consider the pure unitary case \cite{manini00}.
Having a set of initial pure states $\ket{\psi_k},\ k=1,\ldots,n$,
the defining quantity from Eq. (\ref{eq:definitionholonomy1}) can
be written as
\begin{multline}
\label{eq:purestateholonomy}
 W_{j_1}(\tau)W_{j_1}^\dagger (0)W_{j_2}(\tau) W_{j_2}^\dagger(0)
\ldots W_{j_l}(\tau)W_{j_l}^\dagger (0)\\  =  U(\tau) \ket{\psi_{j_1}}
\bra{\phi_{j_1}} B(\tau)\ket{\phi_{j_1}} \bra{\psi_{j_1}}
U(\tau) \ket{\psi_{j_2}} 
\bra{\phi_{j_2}} B(\tau)
\ket{\phi_{j_2}}\\ \bra{\psi_{j_2}} \ldots \bra{\psi_{j_{m-1}}} 
U(\tau) \ket{\psi_{j_l}} \bra{\phi_{j_l}} B(\tau) \ket{\phi_{j_l}}
\bra{\psi_{j_l}},
\end{multline}
where we have used the purified states $\ket{\psi_k}\bra{\phi_k}$.
If $U(t)$ is already parallel transporting the basis states, i.e.,
$\bra{\psi_{j}} U^{\dagger}(t) \dot{U}(t)\ket{\psi_j}=0$, $B(t)$
may be chosen to be the identity and Eq. (\ref{eq:purestateholonomy})
simplifies to
\begin{multline}
\label{eq:purestateholonomy2}
 W_{j_1}(\tau) W_{j_1}^\dagger(0) W_{j_2}(\tau) 
W_{j_2}^\dagger(0) \ldots W_{j_l}(\tau) W_{j_l}^\dagger (0)\\
 =  U(\tau) \ket{\psi_{j_1}}\bra{\psi_{j_1}}
U(\tau) \ket{\psi_{j_2}} 
 \bra{\psi_{j_2}} \ldots
\bra{\psi_{j_{m-1}}} U(\tau) \ket{\psi_{j_l}}\bra{\psi_{j_l}}.
\end{multline}
It is straightforward to write down the off-diagonal phase
factors corresponding to this quantity using $\nu_{\mathcal{C}_{j_1}
\ldots \mathcal{C}_{j_l}}^{(l)} (1)$ to see the equivalence to those
put forward by Manini and Pistolesi \cite{manini00} in accordance with
criterion (a).

What is even more noteworthy is the naturally arising generalisation
of the latter to nonparallel transporting unitarities $U(t)$. A proper
choice of $B(t)$ according to Eq. (\ref{eq:pureparallelity}) yields a
parallel lift and therefore a well-defined invariant of the paths
${\mathcal C}_i$ of the $W_i$'s.

We end this section by noting that the quantity $\nu_{\mathcal{C}_{j_1} 
\ldots \mathcal{C}_{j_l}}^{(l)}(1)$ suggests the alternative ordering
$\mathcal{Y} \equiv W_{j_1}^\dagger (0) W_{j_2}(\tau) W_{j_2}^\dagger (0) 
\ldots W_{j_l}^\dagger (0) W_{j_1}(\tau)$ by shifting $W_{j_1}(\tau)$ 
to the end, since the trace operation is invariant under cyclic
permutations of the constituent operators. As for the gauge invariance
of this alternative ordering, we observe that the global gauge
transformation $W_{j_k}(t) \mapsto W'_{j_k}(t) = W_{j_k}(t) S_{j_k}$
by a $t$ independent partial isometry $S_{j_k}$ yields $\mathcal{Y}
\mapsto S_{j_1}^\dagger \mathcal{Y} S_{j_1}$, i.e., $\mathcal{Y}$ is
dependent on the choice of the initial amplitude. This neither changes
the nodal point structure of $\mathcal{Y}$ nor does it appear when
considering $\nu_{\mathcal{C}_{j_1} \ldots
\mathcal{C}_{j_l}}^{(l)}(1)$.  Thus, both definitions $\mathcal{X}$
and $\mathcal{Y}$ are suitable choices for a proper extension of
Uhlmann's relative phase, though we opt for the former of
Eq. (\ref{eq:definitionholonomy1}) in the course of this work since
there is no need then to refer explicitly to the functional
$\nu_{\mathcal{C}_{j_1}\ldots
\mathcal{C}_{j_l}}^{(l)}(A)$ to obtain a gauge invariant also with
respect to a global gauge.

\section{Comparison to off-diagonal geometric phase for mixed states}
Motivated by the mixed state geometric phase in \cite{sjoqvist00}, the 
present authors have recently introduced a concept of off-diagonal 
geometric phase factors for unitarily evolving mixed states 
\cite{filipp03a}:
\begin{equation}
\gamma_{\rho_{j_1}\rho_{j_2}\ldots\rho_{j_l}}^{(l)}
 \equiv  \Phi \big[ \Tr \big( U(\tau) \sqrt[l]{\rho_{j_1}}
U(\tau) \sqrt[l]{\rho_{j_2}} \ldots
U(\tau) \sqrt[l]{\rho_{j_l}} \big) \big]
\label{eq:offdiaggeom}
\end{equation}
with $\Phi[z]\equiv z/|z|$ for any complex number $z$, the
$\rho_{j_k}$'s only differing by permutations of their eigenstates,
and $U(t)$, $t\in[0,\tau]$, fulfilling parallel transport for each
common eigenstate of the $\rho_{j_k}$'s. For $l=1$ this reduces to the
geometric mixed state phase in \cite{sjoqvist00} that has in general
been shown to be distinct from the trace of the $l=1$
holonomy factor \cite{ericsson03a}. The question is therefore how the
off-diagonal geometric phase definition in \cite{filipp03a}
relates to the off-diagonal generalisation of the Uhlmann phase factor
introduced in the present treatise. Using the same scheme as above to
compensate dynamical effects in the system by an appropriate choice of
unitary operator $B(t) \in \mathcal{B}(\mathcal{H}')$, we get
\begin{eqnarray}
\label{eq:comparisonfull}
 W_{j_1}^\dagger (0) W_{j_2}(\tau) W_{j_2}^\dagger(0) &\ldots&
W_{j_l}(\tau)W_{j_l}^\dagger (0) W_{j_1}(\tau)
\nonumber\\
 & = & \rho_{j_1}^{1/2} (0) U(\tau) \rho_{j_2}^{1/2} (0) B(\tau)
\rho_{j_2}^{1/2} (0) \ldots U(\tau) \rho_{j_l}^{1/2} (0) 
\nonumber\\
 & & \times B(\tau) \rho_{j_l}^{1/2} (0) U(\tau) \rho_{j_1}^{1/2} (0) 
B(\tau) ,
\end{eqnarray}
where the $\rho_{j_k}$'s are those of Eq. (\ref{eq:offdiaggeom}).

In a first guess one could think to obtain a similar form like
$\gamma_{\rho_{j_1}\rho_{j_2}\ldots\rho_{j_l}}^{(l)}$ with a unitarity
$U(t)$ parallel transporting all eigenstates of the $\rho_{j_k}$'s, so
that the $B(t)$ can be chosen to be time-independent. But this procedure
fails since the parallel transport condition behind the
$\gamma_{\rho_{j_1}\rho_{j_2}\ldots\rho_{j_l}}^{(l)}$'s is much weaker
than the parallel transport condition in
Eq. (\ref{eq:paralleltransportoperators}). In the former parallel
transport is required for the state vectors $\ket{\psi_k}$
diagonalising the initial $\rho = \sum_k \lambda_k
\ket{\psi_k}\bra{\psi_k}$, i.e., $\bra{\psi_k}U^{\dagger} (t)
\dot{U} (t) \ket{\psi_k} = 0$, whereas in the latter case
putting $B(t)$ constant amounts to vanishing matrix elements of
$U^{\dagger}(t)\dot{U}(t)$ in the support of $\rho(0)$.
For $\rho_{j_k}$'s having only nonzero eigenvalues this means
that the left-hand side of Eq. (\ref{eq:paralleltransportoperators})
can only vanish for unitarities $U(t)$ that leave all the $\rho_{j_k}$'s
appearing in $\mathcal{X}_{j_1 \ldots j_l}^{(l)} [\mathcal{C}_{j_1}
\ldots \mathcal{C}_{j_l}]$ unaffected or, in other words that the
parallel transport condition is trivially fulfilled for
no evolution at all. For a density operator $\rho$ with 
zero eigenvalues, however, the left-hand side of
Eq. (\ref{eq:paralleltransportoperators}) vanishes, if $U^{\dagger} (t)
\dot{U} (t)$ maps the right support of all $\rho_{j_k}$'s to their 
kernels, i.e. $U^{\dagger} (t) \dot{U} (t):
\textrm{r-supp\,}\rho_{j_k} \mapsto \textrm{Ker\,}\rho_{j_k}$. In this
case $B(t)$ can be set constant also for nonstationary density
operators, as we will see in an example below. Furthermore, the two
approaches are on equal footing in the limit of pure states.

\section{Spin Flip Operation on a Mixture of Bell States}
One explicit example of an evolution that leads to orthogonal initial
and final mixed states is a spin plus phase flip operation applied
to a mixture of Bell states. For the initial state 
\begin{equation}
\label{eq:rho10}
\rho_1(0) =
\frac{1}{1+\varepsilon}\big(\ket{\Psi^-}\bra{\Psi^-} + \varepsilon
\ket{\Psi^+}\bra{\Psi^+}\big) , \  \varepsilon \geq 0,
\end{equation}
we obtain by spin- and phase-flipping the first qubit, i.e., 
$U_{\textrm{sf}}: (\ket{0},\ket{1}) \mapsto (\ket{1},-\ket{0})$ or  
$U_{\textrm{sf}}=\ket{\Phi^+}\bra{\Psi^-}+\ket{\Psi^+}\bra{\Phi^-} - 
\ket{\Psi^-}\bra{\Phi^+}-\ket{\Phi^-}\bra{\Psi^+}$, the final state
\begin{equation}
\label{eq:rho1tau}
\rho_1(\tau) = \frac{1}{1+\varepsilon} \big( \ket{\Phi^+}\bra{\Phi^+} + 
\varepsilon \ket{\Phi^-}\bra{\Phi^-} \big),
\end{equation}
where we have denoted the Bell states by $\ket{\Psi^\pm} =
2^{-1/2}(\ket{01}\pm\ket{10})$ and $\ket{\Phi^\pm} =
2^{-1/2}(\ket{00}\pm\ket{11})$.  A simple implementation of such an
operation is given by the time-independent Hamiltonian $H_s = \sigma_y
\otimes 1_2$ so that the path $\mathcal{C}_1:t \in [0,\tau] \mapsto
\rho_1(t) = U_s(t)\rho_1(0) U_s^\dagger(t)$ with $U_s(t)= e^{-i t
H_s}$ is traced out in state space. Inserting $U_s(t)$ into
Eq. (\ref{eq:paralleltransportoperators}) yields a vanishing left-hand
side, so that we can choose $B_{s1}(t) = 1_1 \otimes 1_2$ to fulfil
the parallel transport condition. For $t=\tau= \pi/2$ we obtain the
amplitude $W_1(\tau)=U_s(\tau) \rho_1^{1/2}(0)=U_{\textrm{sf}} 
\rho_1^{1/2}(0)$ and the $l=1$ holonomy invariant reads 
\begin{eqnarray} 
\mathcal{X}_1^{(1)}[\mathcal{C}_1] & = & 
W_1(\tau)W_1^\dagger(0) = U_{\textrm{sf}} \rho_1(0) 
\nonumber \\ 
 & = & \frac{1}{1+\varepsilon} \big( \ket{\Phi^+}\bra{\Psi^-} - 
\varepsilon \ket{\Phi^-}\bra{\Psi^+} \big) , 
\end{eqnarray} 
which has nonoverlapping right and left support and is therefore 
undefined. In particular, the trace functional $\nu_\mathcal{C} 
(1_1 \otimes 1_2) = \arg \Tr [U_{\textrm{sf}}\rho_1(0)]$, which 
in this special case is the same expression as in 
\cite{sjoqvist00,filipp03a,filipp03b}, vanishes. 

The $l=2$ off-diagonal holonomy invariant can be formed by choosing
the reference state $\rho_2 (0)= \rho_1(\tau)$, which evolves to
$\rho_2(\tau) = \rho_1(0)$ along the path $\mathcal{C}_2: t \mapsto
\rho_2(t) = U_s(t)\rho_2(0)U_s^\dagger(t)$. Again, we can set
$B_{s2}(t) = 1_1 \otimes 1_2$ and obtain $\mathcal{X}_2^{(1)}
[\mathcal{C}_2] = W_2(\tau)W_2^\dagger(0) = U_{\textrm{sf}}\rho_2(0)$,
which also has nonoverlapping left and right support. These
considerations result in
\begin{eqnarray}
\label{eq:l2spinflip1}
\mathcal{X}_{12}^{(2)}[\mathcal{C}_1\mathcal{C}_2] & = & 
W_1(\tau)W_1^\dagger(0)W_2(\tau)W_2^\dagger(0) = 
U_{\textrm{sf}} \rho_1(0) U_{\textrm{sf}} \rho_2(0) 
\nonumber \\ 
 & = & -\frac{1}{(1+\epsilon)^2}\big[\ket{\Phi^+}\bra{\Phi^+} + 
\ket{\Phi^-}\bra{\Phi^-}\big],
\end{eqnarray}
the left and right support of which are overlapping and
$\mathcal{X}_{12}^{(2)}$ is therefore well-defined at this particular 
nodal point of $\mathcal{X}_i^{(1)}[\mathcal{C}_i]$.

The Hamiltonian $H_s$ above is not a unique choice for a spin-flip
implementation, this task can also be performed, e.g., by the
time-dependent Hamiltonian $H_r(t) = [u_z \sigma_z + u_{xy} (\sigma_x
\cos \omega t + \sigma_y \sin \omega t)] \otimes 1_2$ similar to the
Hamiltonian for a resonance spin-flipper (on the first particle)
prevalent in NMR. The unitary time evolution operator corresponding to
$H_r(t)$ can be written as $U_r(t) = U_{\textrm{rot}} U_{\textrm{eff}} = 
e^{-i \omega t \sigma_z/2} e^{-i t H_{\textrm{eff}}} \otimes 1_2$ with 
$H_{\textrm{eff}} = (u_z + \omega/2)\sigma_z + u_{xy} \sigma_x$. By the 
particular choice of the parameters $u_z = -u/2$ and 
$\omega = - 2 u_{xy} = - 2 u_z$, one can verify that for 
$t= \frac{\pi}{\omega}\equiv\tau$ we have the implemented the same
spin-flipping unitary as in the static case,
i.e., $U_r(\pi/\omega) = U_s(\pi/2) = U_{\textrm{sf}}$. Inserting 
$U_r$ on the left-hand side of Eq. (\ref{eq:paralleltransportoperators}) 
we obtain 
\begin{eqnarray}
B_{r1}(t) & = &  
\cos \gamma(t) \Big[ \ket{\Psi^+}\bra{\Psi^+} + 
\ket{\Psi^-} \bra{\Psi^-} \Big] 
\nonumber \\ 
 & & - i \sin \gamma(t) \Big[\ket{\Psi^+}\bra{\Psi^-} + 
\ket{\Psi^-}\bra{\Psi^+} \Big], 
\nonumber \\ 
\gamma(t) & = & \frac{\sqrt{\epsilon}\omega t}{1+\epsilon} .    
\end{eqnarray} 
This gives us the $l=1$ holonomy invariant for the path 
$\widetilde{\mathcal{C}}_1:t \in [0,\tau] \mapsto 
\rho_1(t) = U_r(t)\rho_1(0) U_r^\dagger(t)$ as 
\begin{eqnarray}
\label{eq:hr11}
\mathcal{X}_1^{(1)}[\widetilde{\mathcal{C}}_1] & = & 
W_1(\tau) W_1^\dagger (0) =  
U_{\textrm{sf}} \rho_1^{1/2}(0) B_{r1}(\tau) \rho_1^{1/2}(0) 
\nonumber \\
 & = & \frac{1}{1+\epsilon} \Big[ \cos \gamma (\tau) \left( 
\ket{\Phi^+} \bra{\Psi^-} - \epsilon \ket{\Phi^-} \bra{\Psi^+} \right) 
\nonumber \\ 
 & & + i\sqrt{\epsilon} \sin \gamma (\tau) \left( -\ket{\Phi^+} 
\bra{\Psi^+} + \ket{\Phi^-}\bra{\Psi^-}\right)\Big] ,  
\end{eqnarray}
which has nonoverlapping left and right supports and is therefore 
undefined. Similarly, by again taking $\rho_2(0) = \rho_1(\tau)$ from 
Eq. (\ref{eq:rho1tau}), the $l=1$ holonomy invariant 
associated with the path $\widetilde{\mathcal{C}}_2:t \in [0,\tau] 
\mapsto \rho_2(t) = U_r(t)\rho_2(0) U_r^\dagger(t)$ becomes  
\begin{eqnarray} 
\label{eq:hr12}
\mathcal{X}_2^{(1)}[\widetilde{\mathcal{C}}_2] & = & 
W_2(\tau) W_2^\dagger (0) =  U_{\textrm{sf}} \rho_2^{1/2}(0) 
B_{r2}(\tau) \rho_2^{1/2}(0) 
\nonumber \\
 & = & \frac{1}{1+\epsilon} \Big[ \cos \gamma (\tau) \left( \epsilon 
\ket{\Psi^+} \bra{\Phi^-} - \ket{\Psi^-}\bra{\Phi^+} \right) 
\nonumber \\ 
 & & + i\sqrt{\epsilon}\sin\gamma(\tau)\left(\ket{\Psi^-}\bra{\Phi^-}
- \ket{\Psi^+}\bra{\Phi^+}\right)\Big]  
\end{eqnarray}
with nonoverlapping left and right support. 
 
We may use Eqs. (\ref{eq:hr11}) and (\ref{eq:hr12}) to obtain 
the $l=2$ holonomy invariant 
\begin{eqnarray}
\mathcal{X}_{12}^{(2)}[\widetilde{\mathcal{C}}_1 
\widetilde{\mathcal{C}}_2] & = & 
W_1(\tau)W_1^\dagger(0)W_2(\tau)W_2^\dagger(0) 
\nonumber \\ 
 & = & U_{\textrm{sf}} \rho_1^{1/2}(0) B_{r1}(\tau) \rho_1^{1/2}(0) 
U_{\textrm{sf}} \rho_2^{1/2}(0) B_{r2}(\tau) \rho_2^{1/2}(0) 
\nonumber \\ 
 & = & \frac{1}{(1+\epsilon)^2} \Big[-\big(\cos^2 \gamma(\tau) +
\epsilon\sin^2 \gamma(\tau)\big)\big(\ket{\Phi^+}\bra{\Phi^+} + 
\epsilon\ket{\Phi^-}\bra{\Phi^-}\big)\nonumber\\
 & & + i \sqrt{\epsilon}(1-\epsilon)
\sin\gamma(\tau)\cos\gamma(\tau)\big(\ket{\Phi^+}\bra{\Phi^-} - 
\ket{\Phi^-}\bra{\Phi^+}\big)\Big],
\end{eqnarray}
which has overlapping right and left support and is therefore
well-defined at this particular nodal point of $\mathcal{X}_i^{(1)} 
[\widetilde{\mathcal{C}}_i]$. The difference between 
$\mathcal{X}_{12}^{(2)}[\mathcal{C}_1 \mathcal{C}_2]$ from the 
Hamiltonian $H_s$ and 
$\mathcal{X}_{12}^{(2)}[\widetilde{\mathcal{C}}_1 
\widetilde{\mathcal{C}}_2]$ from $H_r(t)$ reflects the path dependence 
of the off-diagonal holonomy. 

\section{Conclusions}
We have introduced and discussed a concept of off-diagonal quantum
holonomy in connection with the evolution of sets of density
operators. Basically motivated by possible nodal points occurring in
Uhlmann's concept of relative phase
\cite{uhlmann86,uhlmann93,uhlmann96} for some particular paths of
mixed quantum states we have extended the original notion to
off-diagonal quantum holonomy invariants. Utilizing these generalised
quantities the problem of undefined relative Uhlmann phase for initial
and final state with orthogonal supports can be overcome in line with
the introduction of off-diagonal geometric phases for pure states
\cite{manini00}. The definition of the holonomy invariants is
equivalent to the Manini-Pistolesi approach \cite{manini00} in the
pure state limit, moreover it provides us with a natural extension of
the latter to nonparallel-transporting unitary evolutions. When
comparing these holonomy invariants with the off-diagonal mixed state
geometric phases in \cite{filipp03a} we have detected a general
discrepancy for these two approaches related to a fundamental
difference in the treatment of parallel transport of quantum states.
Finally, we have explicitly demonstrated by means of the evolution of
a Bell state mixture the necessity to resort to off-diagonal quantum
holonomies to obtain information about the geometry of state space.

\section*{Acknowledgments}
S.F. acknowledges support from the Austrian Science Foundation,
Project No. F1513. The work by E.S. was supported in part by the 
Swedish Research Council.


\begin{thebibliography}{99}
\bibitem{remark1} In \cite{filipp03a} the initial states are unitarily 
connected by permutation matrices to ensure that the states are 
``in-phase''.
\end{thebibliography}
\end{document}